\documentclass[twocolumn]{article}

\newcommand{\bF}{{\mathbf F}}
\newcommand{\bFt}{{\mathbf F}(t)}
\newcommand{\beq}{\begin{equation}}
\newcommand{\eeq}{\end{equation}}

\newcommand{\zhat}{\hat{\mathbf{z}}}

\usepackage{graphicx}

\title{Controlling qubit transitions during non-adiabatic rapid passage
through quantum interference}

\author{Frank Gaitan \\ Department of Physics \\ Southern Illinois 
University \\ Carbondale, IL 62901-4401}

\begin{document}

\maketitle

\begin{abstract}
In adiabatic rapid passage, the Bloch vector of a qubit is inverted by slowly
inverting an external field to which it is coupled, and along which it is
initially aligned. In {\em non-adiabatic\/} twisted rapid passage, the external
field is allowed to twist around its initial direction with azimuthal angle
$\phi (t)$ at the same time that it is non-adiabatically inverted. For
polynomial twist, $\phi (t)\sim Bt^{n}$. We show that for $n\ge 3$, multiple
qubit resonances can occur during a single inversion of the external
field, producing strong interference effects in the qubit transition 
probability. The character of the interference is controllable through
variation of the twist strength $B$. Both constructive and destructive 
interference are possible, allowing qubit transitions to be greatly enhanced 
or suppressed. Experimental confirmation of these controllable interference 
effects has already occurred. Application of this interference mechanism to 
the construction of fast fault-tolerant quantum controlled-NOT and NOT gates 
is discussed.
\end{abstract}

\section{Introduction}
\label{sec1}

To set the stage for the work to be presented in this paper, we remind the
reader of two fundamental results from the theory of quantum computation:
(1) the existence of universal sets of quantum logic gates; and (2) the
possibility of fault-tolerant quantum computation.

(1) In the quantum circuit model of quantum computation \cite{dd2}, a network
of quantum logic gates is used to implement a desired quantum computation.
An $n$-qubit quantum logic gate is a device that performs a fixed unitary 
transformation $U$ on the state of $n$ qubits.
Just as with classical logic gates, universal sets of quantum logic gates
have been shown to exist so that any quantum computation involving a finite
number of qubits can be carried out using a network composed entirely of 
gates belonging to the universal set \cite{dd2,diV,2qg}. For example,
the 1-qubit Hadamard gate, the 1-qubit phase gate, and the 2-qubit 
controlled-NOT (CNOT) gate form a universal set of gates.

(2) It has been shown that a quantum computation of arbitrary duration using
$n$ qubits can be carried out with arbitrarily small error probability
if all gates used in the computation have an error probability (per gate 
operation) $P$ that falls below a threshold value $P_{a}$ known as the 
accuracy threshold \cite{ath}. A quantum gate is said to operate 
fault-tolerantly if its error probability satisfies $P<P_{a}$. 
Estimates of $P_{a}$ have been made using different error models to
describe the effect of the environment on the quantum gate.
The best known estimate \cite{bch} considered a model in which 
the environment subjects a quantum gate to a classical stochastic process 
that generates independent errors. For this model, $P_{a}\sim 10^{-4}$. Other 
error models have yielded smaller values for $P_{a}$, although the value of 
$10^{-4}$ has become an unofficial benchmark for fault-tolerant operation of 
a quantum gate. Determining how to construct fast quantum gates that also
operate fault-tolerantly ($P< 10^{-4}$) is one of the major technical
challenges facing the quantum computing community. The work reported in this 
paper describes a promising approach for constructing fast fault-tolerant 
quantum NOT and CNOT gates which exploits quantum interference effects
to control qubit transitions. Both of these gates appear often in quantum
algorithms, with the CNOT gate being especially important for quantum
error correction.

To round out this introduction, and to set notation, we briefly summarize
the unitary transformations implemented by quantum NOT and CNOT gates. The 
quantum analogues of the classical bit states $0$ and $1$ are the 1-qubit 
computational basis states (CBS) $|0\rangle$ and $|1\rangle$. Any pair of 
orthonormal basis states can serve as 1-qubit CBS. A quantum NOT gate 
transforms $|0\rangle \longleftrightarrow |1\rangle$ so that its action on the 
1-qubit CBS is $U_{\scriptscriptstyle NOT}\, |i\rangle = |i\oplus 1\rangle$. 
Here $\oplus$ denotes addition modulo~2, and $i=0, 1$.
Since $U_{\scriptscriptstyle NOT}$ is a linear operator, its action on
an arbitrary 1-qubit state follows from its action on the CBS: 
$U_{\scriptscriptstyle NOT}\, \left(\, a|0\rangle + b|1\rangle\,\right) = 
a|1\rangle + b|0\rangle$. The quantum CNOT gate is a 2-qubit gate. The 
2-qubit CBS $|i\, j\rangle$
are obtained by forming all possible tensor products of the 1-qubit CBS:
$|i\, j\rangle = |i\rangle_{c}\otimes |j\rangle_{t}$, with $i,\, j = 0, 1$.
The $c$ ($t$) subscript indicates that the first (second) ket corresponds to 
the {\em control} ({\em target}) qubit. The action of a quantum CNOT gate 
on the 2-qubit CBS
is $U_{\scriptscriptstyle CNOT}\, |i\, j\rangle = |i\rangle_{c}\otimes
|j\oplus i\rangle_{t}$. Thus $U_{\scriptscriptstyle CNOT}$ applies a NOT
operation to the target qubit only when the control qubit has $i=1$. The
action of $U_{\scriptscriptstyle CNOT}$ on an arbitrary 2-qubit state
follows from linearity.

The structure of this paper is as follows. In the next section we examine
3 ways of implementing a quantum NOT gate. The first two are well-known and 
are discussed to show what is possible with familiar technology. The
third approach uses a less familiar form of rapid passage known as 
non-adiabatic twisted rapid passage \cite{fg}. We show that during this 
type of rapid passage the qubit can pass through resonance multiple times 
during a single rapid passage sweep, and that quantum interference effects 
are generated that allow strong 
control of qubit transitions. It is important to note that these quantum 
interference effects have recently been observed using liquid state NMR 
\cite{zwa}. We close Section~\ref{sec2} by showing that a {\em fast 
fault-tolerant\/} quantum NOT gate based on twisted rapid passage is possible 
with {\em existing\/} NMR technology. In section~\ref{sec3} 
we show how a quantum CNOT gate can be implemented using twisted rapid 
passage, and we summarize and make final remarks in Section \ref{sec4}.

\section{Quantum NOT gate: 3 ways}
\label{sec2}

\subsection{Adiabatic rapid passage}
\label{sec2a}

Adiabatic rapid passage (ARP) is a well-known procedure that inverts the 
Bloch vector of a qubit by inverting an external field $\bFt$\ to which it is 
coupled, and along which it was initially aligned. The qubit is coupled to
$\bFt$\ through the Zeeman interaction,
\beq
H(t) = - \mbox{\boldmath$\sigma$}\cdot\bFt \hspace{0.1in} ,
\label{Zeeman}
\eeq
and the external field is inverted such that $\bFt = b\hat{\mathbf{x}} + 
at\hat{\mathbf{z}}$. The instantaneous energy eigenvalues of $H(t)$ are 
$E_{\pm}(t) =\pm\sqrt{b^{2} + (at)^{2}}$, and we denote the instantaneous 
energy eigenstates by $|E_{\pm}(t)\rangle$. From eq.~(\ref{Zeeman}) we see 
that the Bloch vector $\langle\mbox{\boldmath$\sigma$}\rangle_{t}$ is parallel 
(anti-parallel) to $\hat{\bF}(t)$ in the eigenstate $|E_{-}(t)\rangle$ 
($\, |E_{+}(t)\rangle\,$). An avoided crossing occurs at $t=0$
where the energy gap is smallest, and one can show that the qubit is at 
resonance at this crossing (see Section~\ref{sec2c3}). Without loss of 
generality, 
the inversion can be considered to take place during the time interval 
$\left[\, -T/2,\; T/2\,\right]$. For ARP, the inversion time
$T$ is much larger than the inverse Rabi frequency (viz.\ adiabatic), yet 
short compared to the thermal relaxation time $\tau$ (viz.\ rapid). One also 
has that $aT\gg b$ so that $\hat{\bF}(t)$ is effectively aligned with 
$\pm\hat{\mathbf{z}}$ as $t\rightarrow \pm T/2$. The qubit is initially
prepared in an energy eigenstate of the initial Hamiltonian with it's
Bloch vector initially pointing along, say, $-\hat{\mathbf z}$. Then 
$|\psi (-T/2)\rangle = |E_{-}(-T/2)\rangle$. Expanding $|\psi (t)\rangle$ in 
terms of the basis states $|E_{\pm}(t)\rangle$ gives,
\beq
|\psi (t)\rangle = S(t)|E_{-}(t)\rangle + I(t)|E_{+}(t)\rangle 
                          \hspace{0.1in}  .
\label{quantstate}
\eeq
$S(t)$ is the probability amplitude that the qubit will be found
in the $E_{-}$ energy-level at time $t$, and $I(t)$ is the probability
amplitude that a transition has occurred, and that the qubit will be found in 
the $E_{+}$ energy-level at time $t$. The Schrodinger dynamics for ARP can be 
solved exactly for arbitrary values of $a$ and $b$ \cite{L+Z}. The final 
transition probability $P= |I(T/2)|^{2}$ is given by the Landau-Zener 
expression,
\beq
P = \exp\left[\, -\pi /\lambda\,\right] \hspace{0.1in} ,
\label{LZformula}
\eeq
where $\lambda = \hbar a/b^{2}$ (and we are assuming that $a,\; b > 0$).
For ARP, $\lambda\ll 1$ so that the transition probability is exponentially
small. Thus transitions can be safely ignored during ARP, and to an excellent
approximation, we can write:
\beq
|\psi (t)\rangle = |E_{-}(t)\rangle \hspace{0.1in} .
\label{finalstate}
\eeq
Eq.~(\ref{finalstate}) indicates that the qubit Bloch vector remains parallel 
to $\bFt$\ throughout its inversion. Thus ARP causes the qubit Bloch vector to 
be inverted as a consequence of the inversion of the external field $\bFt$.

If we define the CBS so that $|0\rangle = |\sigma_{z} = -1\rangle = 
|E_{-}(-T/2)\rangle$ and $|1\rangle = |\sigma_{z}= +1\rangle =
|E_{-}(+T/2)\rangle$, the previous remarks indicate that ARP causes 
$|0\rangle \longleftrightarrow |1\rangle$. Thus ARP implements a 
quantum NOT operation on the qubit. Note that a transition during ARP means 
that $|E_{\pm}(-T/2)\rangle \rightarrow |E_{\mp}(+T/2)\rangle$, or in terms 
of the CBS: $|i\rangle \rightarrow |i\rangle$. The occurrence of a transition 
during ARP thus corresponds to an error in the quantum NOT operation. The 
Landau-Zener expression (eq.~(\ref{LZformula})) for the transition probability 
$P$ thus gives the error probability (per gate operation) for an ARP quantum 
NOT gate. This error probability can be made arbitrarily small by making the 
inversion take place at a sufficiently adiabatically rate. Thus, ARP can be 
used to implement a fault-tolerant, though adiabatically slow, quantum NOT 
gate.

\subsection{$\pi$--Pulse}
\label{sec2b}

In a $\pi$--pulse, one inverts the qubit Bloch vector through application of 
a pulsed external field $\bFt = F(t)\,\hat{\mathbf{e}}$ whose direction, 
power, and duration are chosen to insure that the Bloch vector undergoes a 
$180^{\circ}$ rotation about $\hat{\mathbf{e}}$. In the usual situation,
the qubit initial state $|\psi (-T/2)\rangle$ is an eigenstate of 
$\sigma_{z}$, and the external field has $\hat{\mathbf{e}} = 
\hat{\mathbf{x}}$. If we write $|\sigma_{z} = -1 \rangle = |\downarrow\rangle$
and $|\sigma_{z} = +1 \rangle = |\uparrow\rangle$, then a $\pi$--pulse maps
$|\downarrow\rangle\longleftrightarrow |\uparrow\rangle$. Defining the
1-qubit CBS as $|0\rangle = |\downarrow\rangle$ and $|1\rangle = 
|\uparrow\rangle$, we see that a $\pi$--pulse implements a quantum NOT gate:
$|0\rangle\longleftrightarrow |1\rangle$. Since the pulse power, duration, and
direction cannot be perfectly controlled, a real $\pi$--pulse will execute an
imperfect quantum NOT operation. Fortunato et.\ al.\ \cite{for} have
worked with non-adiabatic $\pi$--pulses (in NMR) whose error probability 
(per NOT operation) satisfies $P>3\times 10^{-4}$. This is only slightly 
larger than the benchmark value for the accuracy threshold of $10^{-4}$. If 
we denote the pulse amplitude by $F_{1}$, then for $\omega_{1}=\gamma F_{1}
\sim 4000$ Hz ($\gamma = $ gyromagnetic ratio), the pulse duration will be 
$T=\pi /\omega_{1}\sim 1$ msec. Thus the best $\pi$--pulses can implement 
fast, though not quite fault-tolerant, quantum NOT gates.

\subsection{Twisted rapid passage}
\label{sec2c}

Twisted rapid passage (TRP) generalizes ARP in two essential ways: (1) the
adiabatic restriction is relaxed; and (2) the external field $\bFt$\ is 
allowed to twist around its initial direction during the course
of its inversion. Specifically, the time dependence of the external field 
during TRP is: $\bFt = b\cos\phi (t)\,\hat{\mathbf{x}} + b\sin\phi (t)\,
\hat{\mathbf{y}} + at\,\hat{\mathbf{z}}$. In this subsection we show that 
multiple qubit resonances can occur per TRP sweep, and that by varying their 
time separation, quantum interferences effects are produced which allow for 
a direct control over qubit transitions. We then discuss the experimental 
{\em confirmation\/} of this interference
mechanism for controlling qubit transitions, and show how TRP can be
used to implement a {\em fast fault-tolerant\/} quantum NOT gate. A detail
presentation of these results is given in refs.~\cite{fg} and \cite{zwa}. 

\subsubsection{Multiple resonances}
\label{sec2c1}

It proves convenient to transform to the rotating frame in which the
$x$-$y$ component of the external field is instantaneously at rest. This is
accomplished via the unitary transformation $U(t) = \exp\left[\, -(i/2)\phi (t)
\sigma_{z}\,\right]$. The Hamiltonian $\overline{H}(t)$ in this frame is
\beq
\overline{H}(t) = - \mbox{\boldmath{$\sigma$}}\cdot\overline{\bF}(t)
                           \hspace{0.1in} ,
\label{rfHam}
\eeq
and $\overline{\bF}(t) = b\,\hat{\mathbf{x}} + (\, at - \hbar\dot{\phi}
/2\, )\,\zhat$ is the external field as seen in the rotating frame, and
a dot over a symbol represents the time derivative of that symbol. The 
instantaneous energy eigenvalues are $\overline{E}_{\pm}(t) = \pm
\sqrt{b^{2} + (\, at -(\hbar\dot{\phi}/2)\, )^{2}}$. Avoided 
crossings occur when the energy gap is minimum, corresponding to when
\beq
at - \frac{\hbar}{2}\frac{d\phi }{dt} = 0 \hspace{0.1in} .
\label{avdxss}
\eeq
In Section~\ref{sec2c3} we show that qubit resonance occurs at an avoided 
crossing.
For polynomial twist, $\phi_{n}(t) = c_{n}Bt^{n}$, where $B$ is the twist 
strength. The dimensionless constant $c_{n}$ has been introduced to simplify
some of the formulas below. For later convenience we choose $c_{n} =2/n$.
For polynomial twist it is easily checked that eq.~(\ref{avdxss}) always has
the root
\beq
t = 0 \hspace{0.1in} ,
\label{zeroxss}
\eeq
and that for $n\geq 3$, eq.~(\ref{avdxss}) also has the $n-2$ roots,
\beq
t = \left(\mathrm{sgn}\, B\right)^{\frac{1}{(n-2)}}\left(\frac{a}{\hbar |B|}
       \right)^{\frac{1}{(n-2)}} \hspace{0.1in} .
\label{othrxss}
\eeq
All together, eq.~(\ref{avdxss}) has $n-1$ roots, though only the real roots
correspond to qubit resonances. For quadratic twist ($n=2$), only 
eq.~(\ref{zeroxss}) arises. For $n\geq 3$, however, along with the
resonance at $t=0$, real solutions to eq.~(\ref{othrxss}) also occur. The
various possibilities for this situation are summarized in Table~\ref{tbl1}.
\begin{table}[h]
\begin{tabular}{c} \hline\hline\vspace{0.05in}
  (1) \underline{sgn B = +1 \rule{0cm}{0.2in}}  \\
$n$ odd;   resonances at:  $t=0$ 
                             and \vspace{0.05in} 
                 $(a/\hbar B)^{\frac{1}{(n-2)}}$ \\
$n$ even;  resonances at: 
                $t=0$ and \vspace{0.15in}
                   $\pm (a/\hbar B)^{\frac{1}{(n-2)}}$ \\ 
  (2) \underline{sgn B = -1 \rule{0cm}{0.2in}}  \vspace{0.05in}\\
$n$ odd;  resonances at:  $t=0$ 
                          and \vspace{0.05in} 
                  $ -(a/\hbar |B|)^{\frac{1}{(n-2)}}$ \\
$n$ even;  resonance at:  
                $t=0$ \vspace{0.1in} \\ \hline\hline
\end{tabular}
\caption{Classification of regimes under which multiple qubit resonances
occur for polynomial twist with $n \geq 3$\vspace{0.1in}.} 
\label{tbl1}
\end{table}

We see that for polynomial twist with $n \geq 3$, multiple qubit resonances 
always occur per TRP sweep for positive twist strength $B$, while for
negative twist strength, multiple resonances only occur when $n$ is odd.
It is important to note that the time separating the multiple qubit resonances
can be varied through a variation of the twist strength $B$ and/or the 
inversion rate $a$ (see eq.~(\ref{othrxss})).

\subsubsection{Controllable quantum interferences}
\label{sec2c2}

To determine the dynamical impact of TRP we simulated the qubit Schrodinger 
equation numerically in the non-rotating frame. The details of
this simulation are described in ref.~\cite{fg}. The equations 
governing the time evolution of the probability amplitudes $S(t)$ 
and $I(t)$ (see eq.~(\ref{quantstate})) are easily obtained from the 
Schrodinger equation and it is these equations that are numerically
integrated. It proves convenient to re-write these equations 
in dimensionless form. To that end, one introduces the dimensionless time 
$\tau = (a/b)t$, the dimensionless inversion rate $\lambda = \hbar |a|/b^{2}$, 
and the dimensionless twist strength $\eta_{n}$,
\beq
\eta_{n} = \frac{\hbar B}{a}\left(\,\frac{b}{a}\,\right)^{n-2} 
                \hspace{0.1in} .
\label{geneta}
\eeq
From eqs.~(\ref{zeroxss}) and (\ref{othrxss}), the (dimensionless) times
at which the multiple resonances occur are,
\beq
\label{dimzeroxss}
\tau = 0 \hspace{0.1in} ,
\eeq
and
\beq
\tau = \left(\,\mathrm{sgn}\,\eta_{n}\,\right)^{\frac{1}{(n-2)}}\,
          \left[\,\frac{1}{|\eta_{n}|}\,\right]^{\frac{1}{(n-2)}}
            \hspace{0.1in} .
\label{dimothrxss}
\eeq
Only the real solutions of eq.~(\ref{dimothrxss}) correspond to qubit
resonances. Ref.~\cite{fg} examined cubic ($n=3$) and quartic ($n=4$)
TRP in detail. These cases correspond to the simplest examples of odd
and even order twist, respectively, that contain multiple qubit resonances.
Due to space limitations, we only review the results for quartic twist
in this paper. The reader is referred to ref.~\cite{fg} for the cubic TRP
results. For quartic twist $\phi_{4}(t) = (1/2)Bt^{4}$ and $\eta_{4} =
\hbar B b^{2}/a^{3}$. The analysis of Section~\ref{sec2c1} indicates that
qubit resonances will occur at $\tau = 0$ 
and $\tau = \pm 1/\sqrt{\eta_{4}}$ when $\mathrm{sgn}\,\eta_{4} = +1$, and
only at $\tau = 0$ when $\mathrm{sgn}\,\eta_{4} = -1$. The initial condition
for the simulation is $|\psi (-\tau_{0}/2)\rangle = |E_{-}(-\tau_{0}/2)
\rangle$, and $\tau_{0} = (a/b)T$. We will be interested in the transition
probability $P(t) =|I(t)|^{2}$.

\begin{figure}[h]
\centering
\includegraphics[scale=0.3]{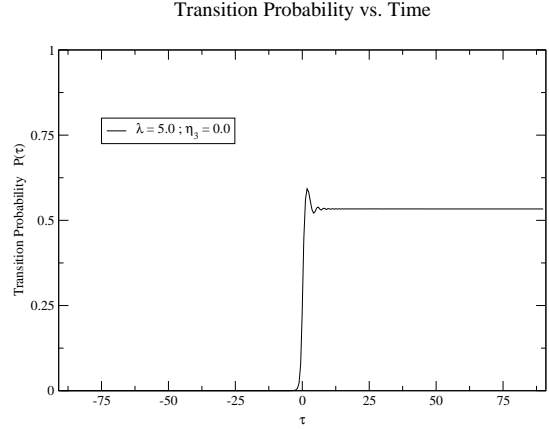}
\caption{\label{fig1}Plot of the transition probability $P(\tau )$ for
twistless non-adiabatic rapid passage with $\lambda = 5.0$ and $\eta = 0$.}
\end{figure}

\begin{figure}[h]
\centering
\includegraphics[scale=0.3]{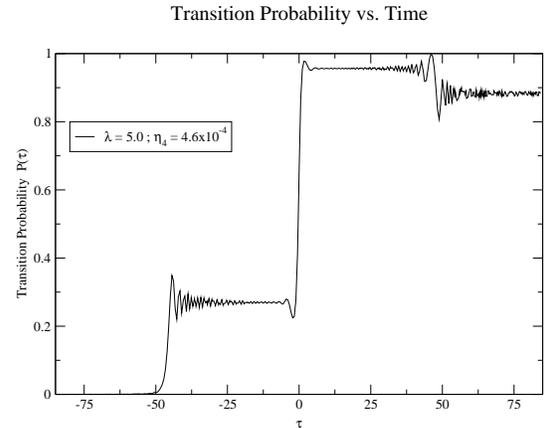}
\caption{\label{fig2}The transition probability $P(\tau )$ for 
non-adiabatic rapid passage with quartic twist with $\lambda = 5.0$ and
$\eta_{4} = 4.6\times 10^{-4}$.}
\end{figure}

\begin{figure}[h]
\centering
\includegraphics[scale=0.3]{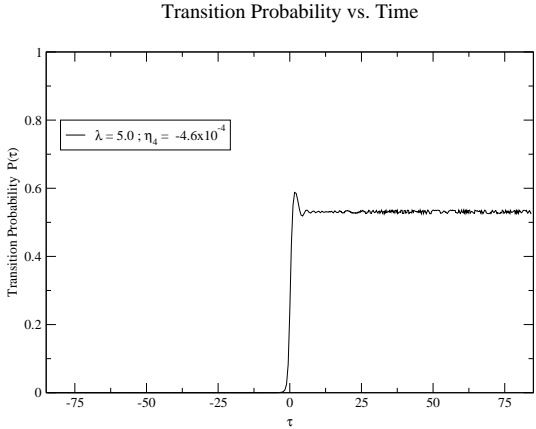}
\caption{\label{fig3}The transition probability $P(\tau )$ for 
non-adiabatic rapid passage with quartic twist with $\lambda = 5.0$ and
$\eta_{4} = -4.6\times 10^{-4}$.}
\end{figure}

For purposes of comparison, Figure~\ref{fig1} shows the transition probability 
$P(\tau )$ vs.\ $\tau$ for {\em twistless\/} rapid passage with 
$\lambda = 5.0$ (non-adiabatic) and $\eta_{4}=0$. 
The final transition probability $P$ at $\tau_{0}/2$ was found to be 
$P = 0.533$. Figure~\ref{fig2} plots the transition probability
$P(\tau )$ for $\lambda = 5.0$ and $\eta_{4}=4.6\times 10^{-4}$. 
The expected qubit resonances at $\tau = 0$ and $\tau = \pm 46.63$ are 
clearly visible. The final transition probability for this case is $P =
0.88$. As we have just seen, twistless rapid passage with $\lambda = 5.0$
has $P=0.533$. Thus the resonances in Figure~\ref{fig2} are constructively
interfering, leading to an enhancement of the qubit transition probability
$P$. Figure~\ref{fig3} shows $P(\tau )$ for quartic twist with $\lambda = 5.0$
and $\eta_{4}=-4.6\times 10^{-4}$. 
This figure clearly shows only one resonance at $\tau = 0$, as expected for 
$\mathrm{sgn}\,\eta_{4}= -1$ (see Table~\ref{tbl1}). The final transition 
probability for this case is $P = 0.533$, which equals the result for 
twistless rapid passage with $\lambda = 5.0$, as one might expect given the 
presence of only one qubit resonance in both cases.

\begin{figure}[h]
\centering
\includegraphics[scale=0.3]{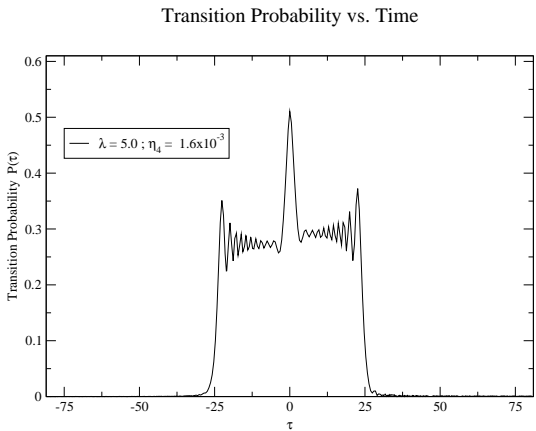}
\caption{\label{fig4}The transition probability $P(\tau )$ for 
non-adiabatic rapid passage with quartic twist with $\lambda = 5.0$ and
$\eta_{4} = 1.6\times 10^{-3}$. Note the slightly reduced vertical scale.}
\end{figure}

Figure~\ref{fig4} plots $P(\tau )$ for $\lambda = 5.0$ and $\eta_{4} =
1.6\times 10^{-3}$. 
The figure clearly shows the expected resonances at $\tau = 0$ and $\tau = 
\pm 25.0$. The final transition probability is $P = 6.93\times 10^{-4}$, 
corresponding to destructive interference relative to twistless rapid passage 
with $\lambda = 5.0$. Note that adding a small amount of quartic twist 
has lowered the final transition probability $P$ by {\em three 
orders-of-magnitude}.
We do not include a plot of $P(\tau )$ for $\lambda = 5.0$ and $\eta_{4} =
-1.6\times 10^{-3}$ as it is similar to Figure~\ref{fig3}, namely one
resonance at $\tau = 0$ and $P=0.533$.

Summarizing these results, we see that: (i) three (one) qubit resonances
(resonance) occur(s) as predicted in Table~\ref{tbl1} when $\mathrm{sgn}\,
\eta_{4} = +1$ ($-1$); (ii) the qubit resonances produce strong interference
effects in the qubit transition probability, with the character of the
interference (constructive or destructive) determined by the time separation
of the resonances; and (iii) the time separation of adjacent qubit resonances 
is given by $\Delta\tau = 1/\sqrt{\eta_{4}}$ ($\mathrm{sgn}\,\eta_{4} = +1$),
and can be controlled through variation of $\eta_{4} = \hbar Bb^{2}/a^{3}$.

\subsubsection{Experimental realization}
\label{sec2c3}

\begin{figure}[h]
\centering\includegraphics[scale=0.9]{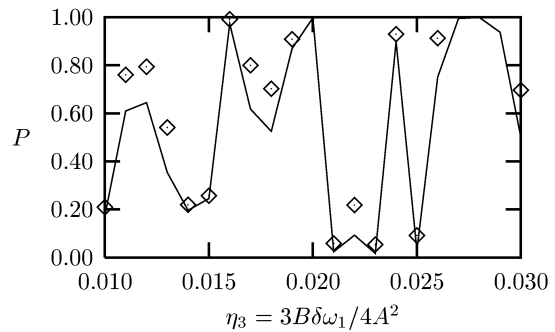}
\caption{\label{fig5}Data and simulation for a cubic sweep profile, as a
function of the dimensionless parameter 
$\eta_{3} = 3B\delta\omega_{1}/4A^{2}$. In the data shown, $A = 50,000$ Hz,
$\delta = 24.39$ Hz, $\omega_{1} = 393$ Hz, and $B$ is calculated from the
target $\eta_{3}$.}
\end{figure}

\begin{figure}[h]
\centering
\includegraphics[scale=0.9]{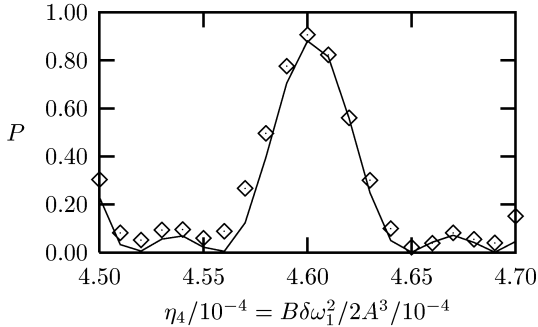}
\caption{\label{fig6}Data and simulation for a quartic sweep profile, 
as a function of the dimensionless parameter 
$\eta_{4} = B\delta\omega_{1}^{2}/2A^{3}$. In the data shown, $A = 50,000$ Hz,
$\delta = 24.39$ Hz, $\omega_{1} = 393$ Hz, and $B$ is calculated from the
target $\eta_{4}$.}
\end{figure}

As pointed out in the Introduction, these multi-resonance induced
quantum interference effects have been experimentally confirmed by
Zwanziger et.\ al.\ using liquid state NMR \cite{zwa}. Both cubic and quartic
twist were experimentally realized. In the experiment a driving rf field is
linearly polarized along the $x$-axis in the lab frame with $F_{x}(t) =
2b\cos\phi_{rf}(t)$. The resonance offset $at$ is produced by linearly
sweeping the detector frequency $\omega_{det}(t)$ through resonance at
the Larmor frequency $\omega_{0}$ such that $\omega_{det}(t) = \omega_{0} +
(2at/\hbar )$. Twist is introduced by sweeping the rf frequency $\omega_{rf}(t)
=\dot{\phi}_{rf}$ through resonance at $\omega_{0}$ in such a way that
$\omega_{rf}(t) = \omega_{det} - \dot{\phi}_{n}$, and $\phi_{n}(t) =
(2/n)Bt^{n}$ is the azimuthal angle for twisted rapid passage introduced in
Section~\ref{sec2c1}. Note that the resonance condition $\omega_{rf}(t) =
\omega_{0}$ is identical to our existence condition for an avoided crossing,
eq.~(\ref{avdxss}). This establishes the promised correspondence between
avoided crossings and qubit resonances. The comparison between experiment and 
theory is given in Figure~\ref{fig5} for cubic twist, and Figure~\ref{fig6} 
for quartic twist.
We see that: (1) variation of the twist strength $\eta_{4}$ clearly causes 
the qubit transition probability $P$ to move between constructive and
destructive interference; and (2) the agreement between theory and experiment
is excellent. We refer the reader to ref.~\cite{zwa} and \cite{gph} for a 
detailed discussion of the experimental parameters, and to the Appendix of 
ref.~\cite{fg} for the translation key that connects our theoretical 
parameters to the experimental parameters of the Zwanziger et.\ al.\ 
experiments.

Before leaving the subject of experimental realization of twisted rapid
passage, two further remarks are in order. First, to insure that all qubits
are inverted when a spread of resonance frequencies occurs, it is 
necessary to require that the frequency sweep cover a large enough interval
that the entire spread of resonance frequencies is included in it. This
gaurantees that all qubits will have passed through resonance by the end of
the frequency sweep. Second, a range of rf field strengths can also be
accommodated so long as $aT/2\gg b_{max}$. This condition insures that the
frequency sweep begins far from resonance for all rf field strengths, and
that transitions will continue to occur only near the avoided crossings. One
therefore anticipates that in this case also, the interference effects will
continue to occur as predicted. For reasonably good samples, magnets, and
rf sources, these constraints can be satisfied, and the interference effects 
presented above should be readily observable. This is in fact what is
found experimentally \cite{zwa}.

\subsubsection{Quantum NOT gate}
\label{sec2c4}
We restrict our attention here to quartic twist, though our discussion 
is readily modified to treat other cases of TRP. As seen in 
Section~\ref{sec2c1}, the external field $\overline{\bF}(t)$ seen in the 
rotating frame has $z$-component $\overline{F}_{z}(t)$ which (written
in terms of dimensionless time $\tau = at/b$) is $\overline{F}_{z}(\tau ) 
= b\tau\, (\, 1
-\eta_{4}\tau^{2}\, )$. Thus the direction of $\overline{\bF}(t)$ approaches
$\mp\zhat$ as $t\rightarrow \pm T/2$. This asymptotic behavior allows us
to implement a quantum NOT gate using quartic twist in a manner that
parallels the ARP approach discussed in Section~\ref{sec2a}. If we initially
prepare the qubit in the $E_{-}$ energy-level, and we define the CBS such that
$|0\rangle = |\sigma_{z} = -1\rangle$ and $|1\rangle = |\sigma_{z} = 
+1\rangle$, then $|\psi (-T/2)\rangle = |E_{-}(-T/2)\rangle = |0\rangle$.
In the absence of transitions, $|\psi (T/2)\rangle = |E_{-}(T/2)\rangle =
|1\rangle$ and TRP with quartic twist thus implements a quantum NOT operation
$|0\rangle\longleftrightarrow |1\rangle$. If a transition occurs, 
$|E_{\pm}(-T/2)\rangle \rightarrow |E_{\mp}(T/2)\rangle$, or $|i\rangle
\rightarrow |i\rangle$, corresponding to an error in the NOT operation. As 
with ARP, the transition probability $P = |I(T/2)|^{2}$ gives the error
probability (per gate operation) of the TRP quantum NOT gate. Table~\ref{tbl2}
gives the transition probabilities for quartic twist pulses for which
$\lambda = 5.0$ and $\eta_{4}$ lies in the interval $[\, 3.95,\; 4.04\,]
\,\times 10^{-3}$.
\begin{table}
\begin{tabular}{cc}\hline\hline
\hspace{0.5in} $\eta_{4}\,(\; \times 10^{-3} \; )$ \hspace{0.3in} & 
                \hspace{0.5in} P \rule[-0.075in]{0cm}{0.225in}
                      \hspace{0.45in} \\ \hline
3.95 & $2.0\times 10^{-2}$ \\
3.96 & $1.3\times 10^{-2}$ \\
3.97 & $6.8\times 10^{-3}$ \\
3.98 & $3.6\times 10^{-3}$ \\
3.99 & $9\times 10^{-4}$ \\
4.00 & $4\times 10^{-5}$ \\
4.01 & $8\times 10^{-4}$ \\
4.02 & $3.9\times 10^{-3}$ \\
4.03 & $1.0\times 10^{-2}$ \\
4.04 & $1.7\times 10^{-2}$ \\ \hline\hline
\end{tabular}
\caption{Transition probabilities for quartic twist with $\lambda = 5.0$ and
$\eta_{4}$ in the range ($3.95$, $4.04$)$\,\times 10^{-3}$.}
\label{tbl2}
\end{table}
The essential thing to notice about Table~\ref{tbl2} is that for $\eta_{4} =
4.00\times 10^{-3}$, the transition/error probability is $P = 4\times 10^{-5}$.
This is {\em less\/} than the benchmark value for fault-tolerant operation of
$10^{-4}\,$! Thus our quartic twist quantum NOT gate is able to operate 
{\em fault-tolerantly\/}. We now show that for pulse parameters which can be 
realized with existing NMR technology, the inversion time for TRP matches that
of a comparable $\pi$-pulse. As shown in ref.~\cite{fg}, the inversion time 
for quartic twist is given by,
\beq
T_{4} = \frac{4A}{\omega_{1}^{2}\lambda} \hspace{0.1in} ,
\label{invtime}
\eeq
where $\omega_{1}$ is related to the amplitude of the NMR rf signal, and
$A$ is related to the TRP inversion rate (see refs.~\cite{fg} and \cite{zwa}).
Current NMR technology can generate $\omega_{1}=4000$Hz and $A = 40,000$ Hz.
With $\lambda = 5.0$, eq.~(\ref{invtime}) gives $T_{4} = 2$ msec. We saw in
Section~\ref{sec2b} that a comparable $\pi$-pulse has an inversion time
$T_{\pi} \sim 1$ msec, and an error probability $P_{\pi}>3\times 10^{-4}$.
Thus a quartic twist quantum NOT gate can match the inversion speed of a
$\pi$-pulse, while delivering an {\em order-of-magnitude\/} smaller error 
probability.
Specifically, quartic twist promises to deliver a {\em fast fault-tolerant\/} 
quantum NOT operation with existing NMR technology. This is a claim
that $\pi$-pulses are currently unable to make. We note that Zwanziger et.\ 
al.\ implemented this particular case of quartic twist, but they were unable 
to resolve a transition probability as small as $P=4\times 10^{-5}$ from
a value of zero \cite{pcm}. Thus, a {\em quantitative\/} test of this 
prediction remains an open experimental challenge.

\section{TRP quantum CNOT gate}
\label{sec3}

\begin{figure}[h]
\centering
\includegraphics[scale=0.4]{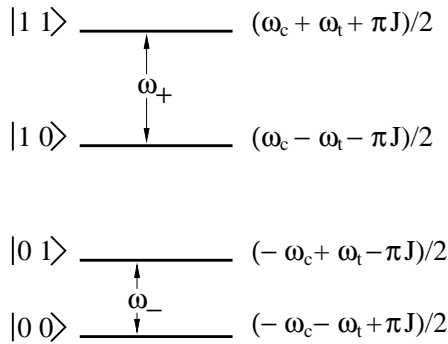}
\caption{\label{fig7}Energy-level structure appropriate for implementing a
quantum CNOT operation using twisted rapid passage. The corresponding energies
(in frequency units) appear to the right of the energy-levels.}
\end{figure}

We now describe a procedure for implementing a quantum CNOT gate using TRP
in the context of liquid state NMR. If the liquid has low viscosity, one
can ignore dipolar coupling between the qubits, and if the remaining
Heisenberg interaction between the qubits is weak compared to the individual
qubit Zeeman energies, it can be well-approximated by an Ising interaction
\cite{g+c}. Under these conditions, the Hamiltonian (in frequency units)
for the control ($c$) and target ($t$) qubits is
\beq
\frac{H_{ct}}{\hbar} = -\omega_{c}I_{z}^{c} - \omega_{t}I_{z}^{t}
                        + 2\pi J\, I_{z}^{c}\, I_{z}^{t} \hspace{0.1in} .
\label{CNOTHam}
\eeq
Here $\omega_{c}$ ($\omega_{t}$) is the resonance frequency of the isolated
control (target) qubit, $J$ is the Ising coupling constant, and $\omega_{c}
> \omega_{t} > \pi J$. We choose the single-qubit CBS to be the eigenstates
of $\sigma_{z}$ with $|0\rangle = |\uparrow\rangle$ and $|1\rangle = 
|\downarrow\rangle$. Then the 2-qubit CBS are $|00\rangle =|\uparrow\uparrow
\rangle$, $|01\rangle = |\uparrow\downarrow\rangle$, $|10\rangle =
|\downarrow\uparrow\rangle$, and $|11\rangle = |\downarrow\downarrow\rangle$,
and they are the eigenstates of $H_{ct}$. The energy levels (in frequency
units) are shown in Figure~\ref{fig7}, where
\beq
\omega_{\pm} = \omega_{t}\pm \pi J \hspace{0.1in} . 
\label{omegapm}
\eeq

Given this energy-level structure, we can implement a quantum CNOT operation
on the two qubits by sweeping through the $\omega_{+}$ resonance using
twisted rapid passage. Refocusing \cite{ich} is used to switch off the
dynamics of the control qubit so that only the target qubit responds to the
TRP pulse. Let $U$ denote the unitary transformation associated with
this procedure. It maps the 2-qubit input state $|\psi_{in}\rangle$ at the
beginning of the procedure to the output state $|\psi_{out}\rangle$:
$|\psi_{out}\rangle = U|\psi_{in}\rangle$. Since the two states $|00\rangle$
and $|01\rangle$ are not resonant, they do not respond to the TRP pulse.
Thus $U|00\rangle = |00\rangle$, and $U|01\rangle = |01\rangle$. On the
other hand, for the $|10\rangle$ and $|11\rangle$ states, the combination
of refocusing and sweeping through the $\omega_{+}$ resonance means that
only the target qubit has its spin inverted. Thus $U|10\rangle = |11\rangle$
and $U|11\rangle = |10\rangle$. This gives the action of $U$ on the 2-qubit
CBS so that we can write out $U$ in the computational basis:
\beq
U = \left(\, \begin{array}{cccc}
                1 & 0 & 0 & 0 \\
                0 & 1 & 0 & 0 \\
                0 & 0 & 0 & 1 \\
                0 & 0 & 1 & 0 
             \end{array} \,\right) \hspace{0.1in} .
\label{UCNOT}
\eeq
The right-hand side of eq.~(\ref{UCNOT}) is recognized as the unitary
transformation implemented by a quantum CNOT gate \cite{ich}, confirming
that our procedure does in fact implement this gate on the two qubits.
Knowing how $U$ acts on the 2-qubit CBS, linearity then determines its 
action on an arbitrary 2-qubit state.

\section{Discussion}
\label{sec4}

In this paper it has been our aim to show that multiple qubit resonances can
occur during a single twisted rapid passage sweep, and that by varying their 
time separation, quantum interference effects are produced which allow for a 
direct control over qubit transitions. This time separation is controlled
through the (dimensionless) twist strength $\eta$ (Section~\ref{sec2c}), and
the resulting interference can be constructive (enhancing transitions) or
destructive (suppressing transitions). These controllable interference effects
are a consequence of the temporal phase coherence of the qubit wave function,
and were experimentally confirmed using liquid state NMR by Zwanziger
et.\ al.\ \cite{zwa}.
Cubic and quartic twist were considered in detail in ref.~\cite{fg} (space 
limitations restricted our discussion here to quartic twist) as they
are the simplest examples, respectively, of odd-order and even-order 
polynomial twist in which these interference effects are expected to occur.
By focusing on these two examples we do not mean to suggest that these
sweep profiles are the best of all possible twisted rapid passage profiles.
A search is currently underway for other profiles that might produce stronger
destructive interference effects. We have also shown how twisted rapid
passage can be used to construct quantum NOT and CNOT gates. It was shown
that parameter values for quartic twist exist that are realizable with
current NMR technology and that will drive a {\em fast fault-tolerant\/}
quantum NOT operation. This feat is currently beyond the 
capabilities of $\pi$-pulse and adiabatic rapid passage quantum NOT gates. 
Other work currently underway includes the following. (1) Development of an 
analytical scheme for approximately calculating the qubit transition 
probability. The aim here being to find trial sweep parameter values that
will yield gate error probabilities $P< 10^{-4}$. These trial values
then serve as the starting point for a more focused numerical search that will
yield the actual parameter values that will drive fast 
fault-tolerant quantum NOT and CNOT operations. (2) We are also working to 
resolve a technical complication associated with the resonance offset that
arises with the quartic twist quantum CNOT gate (see \cite{zwa} for further 
discussion).

\begin{center}
{\bf ACKNOWLEDGMENTS}
\end{center}

I would like to thank T. Howell III for continued support, the National
Science Foundation for support provided through Grant No.\ NSF-PHY-0112335,
and the Army Research Office for support provided through Grant No.\ 
DAAD19-02-1-0051.

\end{document}